\documentclass[twocolumn,pra,amsmath,amssymb,superscriptaddress]{revtex4-1}
\usepackage[utf8]{inputenc}
\usepackage{amsfonts,amsmath,amssymb}
\usepackage{graphicx}
\usepackage{bm} % boldmath
\usepackage{mathrsfs}
\usepackage{subfigure}
\usepackage{textcomp}
\usepackage{microtype}
\usepackage{hyperref}
\usepackage[flushleft]{threeparttable}
\usepackage[per-mode=symbol]{siunitx}
\usepackage[version=3]{mhchem}

\usepackage{booktabs}
\graphicspath{{../Figures/}}

\DeclareSIUnit\intensity{\watt\per\centi\meter\squared}
\DeclareSIUnit\fieldstrength{\volt\per\centi\meter}
\usepackage{xspace}

\DeclareUnicodeCharacter{2009}{\,}

\newcommand{\degree}{\ensuremath{^\circ}}%

\newlength{\figwidth}
\newlength{\figwidthwide}
\let\orgautoref\autoref
\providecommand{\Autoref}{%
  \def\equationautorefname{Equation}%
  \def\figureautorefname{Figure}%
  \def\subfigureautorefname{Figure}%
  \def\tableautorefname{Table}
  \def\sectionautorefname{Section}%
  \orgautoref}
\renewcommand{\autoref}{%
  \def\equationautorefname{Eq.}%
  \def\figureautorefname{Fig.}%
  \def\subfigureautorefname{Fig.}%
  \def\sectionautorefname{Sec.}%
  \orgautoref}

\usepackage{color}
\usepackage[normalem]{ulem}
\definecolor{darkgreen}{rgb}{0.0,0.7,0.0}

\begin{document}

\title{Laser-induced Coulomb explosion of the LiI molecule and of its dimer} %Title of paper

\author{Henrik H. Kristensen}
\affiliation{Department of Chemistry, Aarhus University, Langelandsgade 140, DK-8000 Aarhus C, Denmark}
\author{Emil Hansen}
\affiliation{Department of Physics, Aarhus University, Ny Munkegade 120, DK-8000 Aarhus C, Denmark}
\author{Jeppe K. Christensen}
\affiliation{Department of Chemistry, Aarhus University, Langelandsgade 140, DK-8000 Aarhus C, Denmark}
\author{Simon H. Albrechtsen}
\author{Frank Jensen}
\author{Henrik Stapelfeldt}
\email[]{henriks@chem.au.dk}
%\thanks{}
\affiliation{Department of Chemistry, Aarhus University, Langelandsgade 140, DK-8000 Aarhus C, Denmark}

\date{\today}

\begin{abstract}
A gas-phase sample consisting of lithium iodide, \ce{LiI}, molecules and their dimer \ce{(LiI)2}, are Coulomb exploded by an intense 25~femtosecond laser pulse. In the case of \ce{LiI}, we focus on the double ionization that creates a pair of \ce{Li^+} and \ce{I^+} recoil ions. From the kinetic energy distribution of the \ce{Li^+} ions, extracted using coincidence filtering, we determine the distribution of internuclear distances $P(R)$ via the ground state potential curve of \ce{LiI^{2+}} obtained from an \emph{ab initio} calculation that accounts for non-Coulombic effects.  We find that the center of $P(R)$ is close to the expected internuclear separation based on the three vibrational states of \ce{LiI} populated, whereas the width of $P(R)$ exceeds the theoretical value by $\sim$  52~\%.  We discuss if fragmentation via excited \ce{LiI^{2+}} potential curves affects the determination of $P(R)$. In the case of the dimer, \ce{(LiI)2}, we observe kinetic energies and relative emission directions of \ce{Li^+}, \ce{I^+}, and \ce{I^{2+}} recoil ions consistent with Coulomb explosion of the parallelogram-shaped dimer after removing up to six electrons by the laser pulse.

\end{abstract}

\maketitle

\section{Introduction}\label{sec:intro}

When an intense laser pulse multiply ionizes a molecule, the multiply-charged molecular cation may break apart into ionic fragments. Their momentum vectors are determined by the structure and spatial orientation of the molecule at the arrival time of the laser pulse. The fragmentation process is termed laser-induced Coulomb explosion~\cite{yatsuhashi_multiple_2018,schouder_laser-induced_2022,li_ultrafast_2022} and during its almost 40 years of studies, two main applications have emerged. The first is to use the emission directions of the fragment ions as a measure for how the molecules are turned in space. This has been explored extensively in studies where moderately intense laser pulses align or orient molecules – mostly in the gas phase~\cite{stapelfeldt_colloquium:_2003,ohshima_coherent_2010,fleischer_molecular_2012,koch_quantum_2019}, but also when located in helium nanodroplets~\cite{pentlehner_impulsive_2013,chatterley_long-lasting_2019,qiang_femtosecond_2022,kranabetter_nonadiabatic_2023}. The second application is to use the momentum vectors of the recoil ions to extract information about the molecular structure, i.e., the binding angles and bond distances of molecules – either in static structures~\cite{hasegawa_coincidence_2001,pitzer_direct_2013,voigtsberger_imaging_2014,slater_covariance_2014,kunitski_observation_2015,boll_x-ray_2022, yu_capturing_2022,kranabetter_structure_2024} or as they undergo nuclear dynamics such as dissociation~\cite{stapelfeldt_wave_1995,amini_photodissociation_2018,lam_simultaneous_2025,li_imaging_2025,wang_conformer-sensitive_2025}, vibration~\cite{ergler_spatiotemporal_2006,rudenko_real-time_2006,jyde_time-resolved_2024,lam_simultaneous_2025}, isomerization~\cite{livshits_time-resolving_2020,livshits_symmetry-breaking_2024} or roaming~\cite{ibrahim_tabletop_2014,endo_capturing_2020,mishra_direct_2024}. These applications require that the laser pulse is sufficiently short that nuclear motion during the pulse is minimized.

Determination of bond distances relies on measurements of the kinetic energy of the recoil ions, and requires that there is a one-to-one correspondence between the bond distance and the kinetic energy. Most studies have focused on dimers, notably diatomic molecules. In such cases, a diatomic molecule \ce{AB} is, e.g., doubly ionized leading to fragmentation into an (\ce{A^+}, \ce{B^+}) ion pair. A common assumption has been that \ce{A^+} and \ce{B^+} move apart due to pure Coulomb repulsion, meaning that their potential energy immediately after double ionization is:
\begin{equation}
\label{eq:Coulomb}
\begin{aligned}
E_\text{pot} = E_\text{Coul} = \dfrac{14.4~\text{eV}}{R\text{[\AA]}},
\end{aligned}
\end{equation}
where $R$ is the internuclear separation at the time when the laser pulse arrives. Energy and momentum conservation imply that the final kinetic energy of \ce{A^+} (in the dication frame) is
\begin{equation}\label{eq:Ekin}
E_\text{kin}(A^+) = \dfrac{m_B}{m_A + m_B}E_\text{Coul}
\end{equation}
and similar for $E_\text{kin}(B^+)$. Therefore, $E_\text{Coul}$ is obtained by measuring either $E_\text{kin}(A^+)$ or $E_\text{kin}(B^+)$ and then $R$ can be determined through \autoref{eq:Coulomb}. This principle has been applied in several works and even used to determine the distribution of $R$ of static molecules~\cite{zeller_imaging_2016,kristensen_laser-induced_2023} as well as of dissociating or vibrating~\cite{ergler_spatiotemporal_2006,shi_tracking_2023,jyde_time-resolved_2024} molecules by measuring the (time-dependent) distribution of $E_\text{kin}$.

Describing the potential energy of a multiple ionized dimer as Coulombic, \autoref{eq:Coulomb}, is however only valid in certain situations. These include doubly-ionized \ce{H2} and \ce{D2}~\cite{chelkowski_femtosecond_1999,ergler_spatiotemporal_2006,rudenko_real-time_2006} and, to a very good approximation, doubly-ionized \ce{He2}~\cite{zeller_imaging_2016} and other dimers where all valence electrons are stripped off rapidly. The latter case was demonstrated recently for dimers of alkali atoms, residing at the surface of helium nanodroplets, where only two electrons need to be removed~\cite{kristensen_laser-induced_2023,albrechtsen_laser-induced_2024}. The approximation is only good provided $R$ is not too small, see, e.g., Ref. \cite{kristensen_quantum-state-sensitive_2022}. For most other dimers of atoms or molecules, all valence electrons are typically not stripped off in experiments, although intense x-ray pulses from free-electron lasers have the potential to do this~\cite{rudenko_femtosecond_2017,takanashi_ultrafast_2017}. The residual valence electrons lead to non-Coulombic effects, which result in several, often many, electronic states of the doubly-ionized or multiply-ionized dimer~\cite{wright_dissociation_1999,schouder_laser-induced_2020}, some of which may be metastable. Thereby, the one-to-one correspondence between the kinetic energies of the fragment ions and the interatomic or intermolecular bond distances is lost, i.e., accurate structure determination from the experimental observables may not be possible. This was illustrated explicitly for the \ce{CS2} dimer~\cite{schouder_laser-induced_2020} and is expected to have also been the case in previous studies on other dimer systems~\cite{wu_structures_2012,wu_communication:_2014}.

Here we explore femtosecond (fs) laser-induced Coulomb explosion of a diatomic molecule, lithium iodide, LiI, under conditions where only two out of the total six valence electrons are removed, i.e., the Coulomb fragmentation process: \ce{LiI^{2+}}~$\rightarrow$~\ce{Li^+}~+~\ce{I^+}. Based on the considerations above, one would at first expect that the multiple potential energy curves of \ce{LiI^{2+}} resulting from the lowest molecular orbitals, see \autoref{fig:LiI_potentials}(b) and \autoref{sec:principle}, prevent an accurate determination of the distribution of internuclear distances $P(R)$. However, we find that $P(R)$ obtained from the distribution of kinetic energies of the \ce{Li^+} recoil ions via the \ce{LiI^{2+}} potential energy curve calculated for the lowest-lying state of the molecular dication agrees well with the theoretical $P(R)$. The latter is given by the weighted sum of the internuclear wave functions for the three vibrational states of \ce{LiI} populated at the temperature of the molecular sample.

It turns out that the molecular sample contains \ce{LiI} dimers in addition to the \ce{LiI} monomers. Upon irradiation of the sample with the intense laser pulse, the dimers are also multiply-ionized and ionic fragments from the ensuing Coulomb explosion processes show up as distinct peaks in the kinetic energy distributions of \ce{Li^+}, \ce{I^+}, and \ce{I^{2+}}. Thus, although the focus of the work here is on Coulomb explosion imaging of the \ce{LiI} monomer, it is necessary to also identify the ion signals from the dimers. The second part of the paper is devoted to that and shows the capability of laser-induced Coulomb explosion in combination with covariance analysis of the momentum and emission direction of the fragment ions to identify and distinguish the constituents of a mixed sample~\cite{burt_communication:_2018,pickering_femtosecond_2018,pathak_differentiating_2020,yu_determining_2021,lam_differentiating_2024,albrechtsen_laser-induced_2024}.

\section{Principle of Coulomb explosion imaging of L\lowercase{i}I}\label{sec:principle}

\begin{figure}[t!]
\includegraphics[width = 8.6cm]{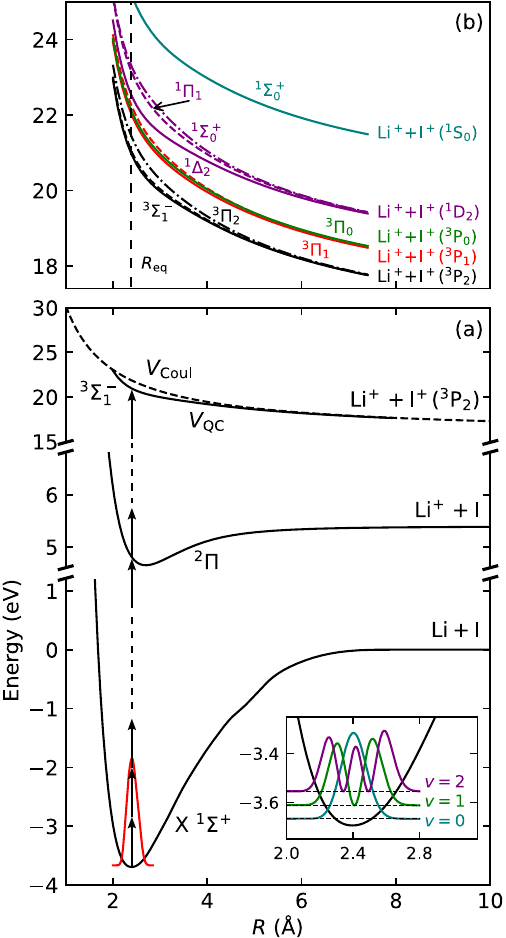}
\caption{(a) Potential energy curves for the ground state X~$^1\Sigma^+$ of \ce{LiI}, ground state $^2\Pi$ of \ce{LiI+}~\cite{schmidt_predissociation_2015} and for the lowest-lying electronic state of \ce{LiI^{2+}}, $V_\text{QC}$. The Coulomb potential is shown as a dashed curve. The inset shows the square of the vibrational wave function for $v$~=~0, 1, and 2, and the red shape shows the Boltzmann weighted average of the three states. The vertical black arrows illustrate the multiphoton absorption causing sequential double ionzation of \ce{LiI}. (b) Potential curves for the 10 states of \ce{LiI^{2+}} arising from the lowest-lying orbitals -- see text. Note there are two close-lying $^3\Sigma_1^-$ and $^3\Pi_0$ curves but only one term symbol is indicated in each case. The \ce{I+} terms are given to the right of the potential curves. }
\label{fig:LiI_potentials}
\end{figure}

\Autoref{fig:LiI_potentials} shows an energy level diagram of \ce{LiI} and \ce{LiI^{2+}}. Initially, the LiI molecules are in the electronic ground state, X $^1\Sigma^+$. Assuming thermal equilibrium of the $T=430\degree$~C gas-phase sample, LiI is essentially only populated in the three lowest-lying vibrational states with populations $N_v$, according to the Boltzmann distribution, $v$ denoting the vibrational quantum number: $N_0$ = 64~\%, $N_1$ = 24~\%, $N_2$ = 9~\% ($N_3$ = 3~\% is neglected). The distributions of internuclear distances, $P_v(R)$, i.e., the square of the vibrational wave function, for each of the three vibrational states, are shown in the inset on \autoref{fig:LiI_potentials}(a). The three $P_v(R)$ were obtained by solving the one-dimensional stationary vibrational Schrödinger equation for the X state. The purpose of applying Coulomb explosion imaging is to determine the internuclear distance distribution. Since the experiment represents an average over all molecules in the sample, we expect to probe the weighted internuclear distribution over the three vibrational states, i.e., $P_\text{theo}(R) = \sum_{v=0}^{2}N_v P_v(R)$, rather than the individual $P_v(R)$. In \autoref{fig:LiI_potentials}(a) $P_\text{theo}(R)$ is illustrated in red near the bottom of the potential curve for \ce{LiI}.

The triggering event in the laser-induced Coulomb explosion imaging is double ionization, through multiphoton absorption, of the \ce{LiI} molecules, see \autoref{fig:LiI_potentials}(a). The double ionization, expected to occur sequentially, projects the vibrational wave function of \ce{LiI} onto a potential curve of \ce{LiI^{2+}}, possibly subject to an $R$-dependence of the ionization process, see \autoref{sec:results:internuclear}. The repulsive character of the potential causes \ce{LiI^{2+}} to break up into a pair of \ce{Li+} and \ce{I+} fragment ions.  During the electrostatic repulsion, the potential energy of \ce{LiI^{2+}}, acquired just after the double ionization, is converted to kinetic energy of the \ce{Li+} and \ce{I+} ions. The simplest approach to determining $P(R)$ is then to apply \autoref{eq:Coulomb} to the distribution of the total kinetic energy release, $P(E_\text{kin}^{tot})$, obtained from one of the fragments, here \ce{Li+} as described below. While this approach was shown to work well for alkali dimers~\cite{kristensen_laser-induced_2023,albrechtsen_laser-induced_2024}, we do not expect it to yield accurate results here due to non-Coulombic effects in the \ce{LiI^{2+}} ion. Therefore, we calculate the potential for the ground state of \ce{LiI^{2+}}, denoted $V_\text{QC}$, and use that instead of $V_\text{Coul}$ to determine $P(R)$. $V_\text{QC}$ was calculated at the CCSD(T) level with 28 electrons correlated for the lowest triplet state $^3\mathrm{\Sigma}_1^-$, using the aug-cc-pwCVTZ and aug-cc-pwCVQZ basis sets and employing an L$^{-3}$ extrapolation~\cite{peterson_molecular_2010,g16,halkier_basis-set_1999}. \Autoref{fig:LiI_potentials}(a) shows that in the pertinent region around the equilibrium distance $R_\text{eq}$ of \ce{LiI}, $V_\text{QC}$  deviates significantly from the Coulomb curve, $V_\text{Coul} = E_\text{Coul} +$~{\it IE}(\ce{I})~+~{\it IE}(\ce{Li}), where the latter two terms are the ionization energies of the \ce{I} and \ce{Li} atoms, respectively. In addition, we have calculated all potential energy curves arising from distributing six electrons in the four valence orbitals and including spin-orbit interactions, with details provided in the appendix, \autoref{sec:Appendix}. The calculation shows that there is a total of 10 energetically-different states. The corresponding potential curves are displayed in \autoref{fig:LiI_potentials}(b). In \autoref{sec:results:internuclear} we discuss possible population of the excited states of \ce{LiI^{2+}} by the laser pulse and, if that happens, how fragmentation on the excited potential curves affect the determination of $P(R)$.

\section{Experimental setup}\label{sec:setup}

The experiment was conducted on an apparatus described in detail previously~\cite{kristensen_laser-induced_2023}, so only a few key points are presented here. An effusive beam containing \ce{LiI} monomers and \ce{LiI} dimers, \ce{(LiI)2}, propagating in the $z$-direction, is created by heating a sample of \ce{LiI} to 430~$\degree$C in an oven inside a vacuum chamber. At this temperature, the partial pressure of the monomer is $\sim3\times10^{-5}$~mbar, $\sim6\times10^{-5}$~mbar of the dimer and $\sim6\times10^{-7}$~mbar of the trimer~\cite{bencze_evaporation_1998}, i.e., the amount of trimers is negligible. The gas-phase \ce{LiI} and \ce{(LiI)2} molecules exit the oven via a 3-mm-diameter hole and travel into a velocity map imaging (VMI) spectrometer at the center of which the effusive beam is crossed by a focused, pulsed laser beam that propagates in the $y$-direction ($\lambda$ = 800~nm, $\tau_\text{FWHM}$ = 25~fs, $I_0$ = $4.7 \times 10^{14}$~W/cm$^2$). The laser pulses, originating from an amplified Ti-Sapphire laser system (Solstice Ace, Spectra-Physics, 1-kHz, $\tau_\text{FWHM}$ = 50~fs), are spectrally broadened in an argon-filled hollow-core capillary tube and subsequently temporally compressed to yield the 25-fs-long pulses used for the measurements. The pulse duration is characterized with an interferometric autocorrelator. The VMI spectrometer projects the ions created by the laser pulses onto a position-sensitive detector with an energy resolution of 100~meV in the explored energy region. The detector is monitored by a TPX3CAM that allows simultaneous recording of all produced ion species~\cite{fisher-levine_timepixcam_2016,zhao_coincidence_2017,nomerotski_imaging_2019}. With the TPX3CAM we can obtain both the two-dimensional momentum and the time-of-flight (ToF) of each individual ion, enabling detailed coincidence analysis and filtering. The direction of the electric field from the VMI spectrometer is along the $x$-axis, and the laser beam is linearly polarized parallel (along $z$) to the $y$-$z$ detector plane.

\section{Results and Discussion}\label{sec:results}

\subsection{LiI monomer}

\subsubsection{Identification of \ce{LiI}: Ion images, covariance maps and kinetic energy distributions}

\begin{figure*}
\includegraphics[width = 17.2cm]{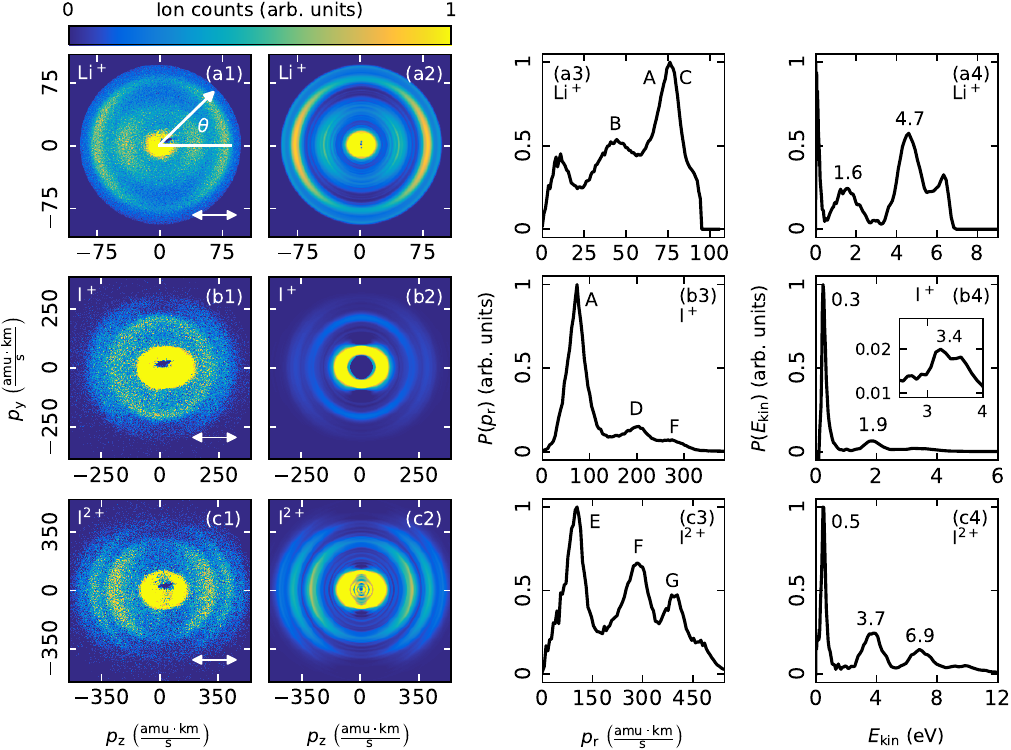}
\caption{(a1)--(c1) 2D momentum images for (a1) \ce{Li+}, (b1) \ce{I+}, and (c1) \ce{I^{2+}}, scaled individually and with saturated colors for improved visual contrast. The white arrows in the bottom right corners show the polarization axis of the laser pulse. A linear colormap scale has been used for the plots. (a2)--(c2) Slices through the center of the reconstructed 3D momentum distributions, obtained by Abel inversion of the images in (a1)--(c1). (a3)--(c3) Radial momentum distributions $P(p_\text{r})$ in the detector plane. The peaks are labeled A--G, referring to different Coulomb fragmentation channels. The peaks associated with channel A and C are overlapping. (a4)--(d4) Kinetic energy distributions $P(E_\text{kin})$. The central positions of the peaks are annotated next to them. A zoomed-in view of the \ce{I+} peak associated with channel F is shown in the inset in (b4).}
\label{fig:images_distributions}
\end{figure*}

\Autoref{fig:images_distributions}(a1) shows the two-dimensional (2D) momentum image obtained for \ce{Li+} ions, obtained by selecting the ion signal with the ToF associated with \ce{^7Li+} in the recorded data because \ce{^7Li} is the most abundant isotope, (92.2 \%). The image has several distinct features at different radii. They stand out as peaks in the radial momentum distribution in the detector plane $P(p_\text{r})$, \autoref{fig:images_distributions}(a3), obtained by angular integration of the image. Next, the distribution of kinetic energies, $P(E_\text{kin})$, is determined. This is done by first Abel inverting the 2D momentum image using the POP (polar onion peeling) algorithm~\cite{roberts_toward_2009} yielding the three-dimensional (3D) momentum image. \Autoref{fig:images_distributions}(a2) shows a slice through the center of this 3D image. From the 3D image, we retrieve the momentum distribution and then convert it to $P(E_\text{kin})$, \autoref{fig:images_distributions}(a4), by applying an energy calibration for the VMI spectrometer and the standard Jacobian transformation~\cite{stewart_calculus_2006}. The central positions of the two main peaks are indicated on the figure~\footnote{The peak close to 0~eV in the kinetic energy distribution for \ce{Li+}, see Fig.~\ref{fig:images_distributions}(a1), is, presumably, due to dissociative ionization of \ce{LiI} into \ce{Li+} and \ce{I}, and is not of interest in this study.}. Notably, the position of the second peak, 4.7~eV, is close to the energy of 4.86~eV that \ce{Li+} is expected to receive during Coulomb explosion of \ce{LiI^{2+}} into \ce{Li+} and \ce{I+} via $V_\text{QC}$ when starting from  $R_\text{eq}$~=~2.39~{\AA} of LiI. The 2D momentum ion image for the \ce{I+} ions is presented in \autoref{fig:images_distributions}(b1). Similarly to the \ce{Li+} image, it contains multiple features, which appear as peaks in $P(p_\text{r})$ and in $P(E_\text{kin})$, see \autoref{fig:images_distributions}(b3)--(b4). The peaks in $P(E_\text{kin})$ are centered at 0.3~eV, 1.9~eV and 3.4~eV. Here, the first peak matches the 0.27~eV of kinetic energy that \ce{I+} is expected to obtain from Coulomb explosion of \ce{LiI} into \ce{Li+} and \ce{I+} via $V_\text{QC}$ starting from $R_\text{eq}$.

\begin{figure}
\includegraphics[width = 8.6cm]{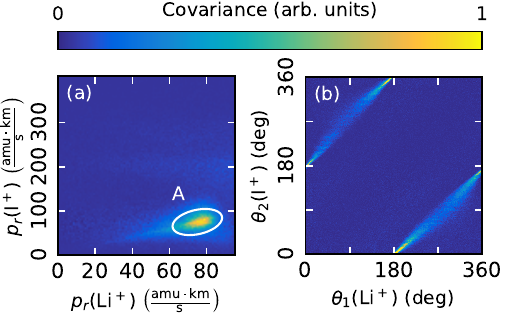}
\caption{(a) Covariance map of the radial momentum distributions for \ce{Li+} and \ce{I+}. (b) Covariance map for the angular distributions of \ce{Li+} with 30 $<$ $p_\text{r}(\text{Li}^+)$ $<$ 95~$\text{amu}\cdot\text{km} / \text{s}$ and \ce{I+} with 30 $<$ $p_\text{r}(\text{I}^+)$ $<$ 95~$\text{amu}\cdot\text{km} / \text{s}$. The colormap scale is linear, and the colors in (b) have been saturated to increase the visual contrast.}
\label{fig:Li_I_covariance}
\end{figure}

To corroborate that the peaks at 4.7~eV for \ce{Li+} and at 0.3~eV for \ce{I+} contain ions originating from Coulomb explosion of \ce{LiI}, we determined the covariance map for the radial momentum distributions of the \ce{Li+} and \ce{I+} ions~\cite{frasinski_covariance_2016,vallance_covariance-map_2021}. The covariance map is shown in \autoref{fig:Li_I_covariance}(a). The single, elongated covariance island observed shows a correlation between \ce{Li+} and \ce{I+} ions with radial momenta of around 75~$\text{amu}\cdot\text{km} / \text{s}$, corresponding to the ions in the kinetic energy distribution peaks at 4.7~eV for \ce{Li+} and 0.3~eV for \ce{I+}. Such even momentum sharing is consistent with two-body breakup of a diatomic molecule. We designate this fragmentation channel as channel A, and have annotated the peaks in $P(p_\text{r})$, \autoref{fig:images_distributions}(a3)--(b3), accordingly.  As an additional test, we also determined the covariance map for the angular distributions of the \ce{Li+} and \ce{I+} ions in channel A, that is, with radial momenta between 30 and 95~$\text{amu}\cdot\text{km} / \text{s}$, see \autoref{fig:Li_I_covariance}(b). In the angular covariance map, there are two diagonal lines centered at $\theta_2 = \theta_1 \pm 180\degree$ ($\theta$ is defined in \autoref{fig:images_distributions}(a1)) indicating that the \ce{Li+} and \ce{I+} ions in channel A are ejected back to back, consistent with a two-body breakup of \ce{LiI^{2+}}. Intensity maxima at $0\degree/360\degree$ and $180\degree$ reflects an anisotropy in the angular distribution of the two ion species, directly visible in both the 2D momentum images and the slices through the center of the corresponding 3D images, \autoref{fig:images_distributions}(a1)--(a2), (b1)--(b2). We ascribe the anisotropy to an alignment-dependent double ionization probability peaking when the internuclear axis of a \ce{LiI} molecule is parallel to the polarization direction of the laser pulse. This phenomenon, observed for many other molecules, is often referred to as geometric alignment~\cite{posthumus_dynamics_2004}. We note that there is an angle-dependent width of the covariance lines. This is due to the velocity spread of the \ce{LiI} molecules in the effusive beam. It could be avoided if the \ce{LiI} molecules instead were delivered in a supersonic beam with a narrow velocity distribution, in which case the covariance lines would get a width of $\sim 1\degree$ as observed previously for Coulomb explosion of e.g. \ce{I2} molecules~\cite{shepperson_strongly_2017}.

\subsubsection{Determination of the distribution of internuclear distances}\label{sec:results:internuclear}

Now we show that the distribution of internuclear separations $P(R)$ for \ce{LiI} can be retrieved from the peak centered at 4.7~eV in the kinetic energy distribution of the \ce{Li+} ions, despite the fact that the peak overlaps with ions from a channel with higher kinetic energies, see \autoref{fig:images_distributions}(a4). To filter out the 4.7~eV peak and thereby extract the \ce{Li+} ions that originate from the Coulomb explosion channel, we retain only the \ce{Li+} (\ce{I+}) ions that have a matching \ce{I+} (\ce{Li+}) partner, recorded within the same laser shot, fulfilling the following criteria for their momenta and ejection directions:
\begin{align}
\left| p_\text{r}(\text{Li}^+) - p_\text{r}(\text{I}^+) \right| < 20 \frac{\text{amu}\cdot\text{km}}{\text{s}} \label{eq:filter_rad}\\
-15\degree < \left| \theta_1(\text{Li}^+)-\theta_2(\text{I}^+) \right| - 180\degree < 15\degree. \label{eq:filter_ang}
\end{align}
Such coincident filtering was previously applied to separate ions from Coulomb explosion of heteronuclear alkali dimers in a mixed sample~\cite{albrechtsen_laser-induced_2024}.

\begin{figure}
\includegraphics[width = 8.6cm]{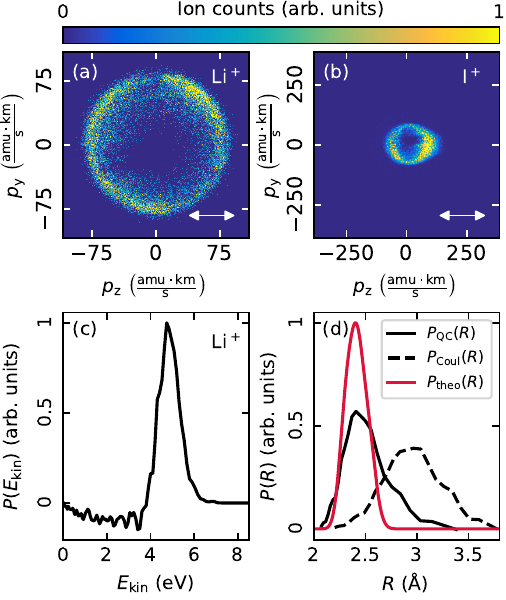}
\caption{Coincidence-filtered \ce{Li+} (a) and \ce{I+} (b) ion images for channel A, plotted with a linear color scale. (c) Kinetic energy distribution of the \ce{Li+} ions in (a). (d) Distributions of internuclear distances for \ce{LiI}, extracted from the kinetic energy distribution in (a) using the \ce{LiI^{2+}} potential (black, solid curve) or the Coulomb potential (black, dashed curve). The theoretical reference, $P_\text{theo}(R)$, is also shown (red curve).}
\label{fig:wavefunctions}
\end{figure}

\Autoref{fig:wavefunctions}(a)--(b) shows the 2D momentum images for the \ce{Li+} and \ce{I+} ions that passed the filter defined by \autoref{eq:filter_rad} and \autoref{eq:filter_ang}. Now, only one ring is visible in each image. These rings with $p_\text{r} \sim 75$~$\text{amu}\cdot\text{km} / \text{s}$ correspond to channel A, i.e, the ions produced from Coulomb explosion of \ce{LiI^{2+}} into \ce{Li+} and \ce{I+}. This is the expected and desired result of the coincidence filtering. The weaker signal in the upper left part and at the bottom of the \ce{I+} image is caused by the reduced sensitivity of the MCP detector in the region near the center. Consequently, a weak signal $180\degree$ away, i.e., in the lower right part and at the top, is introduced in the coincidence-filtered \ce{Li+} image. Its effect on the radial distribution and the kinetic energy distribution of the \ce{Li+} ions, which is our interest here, is, however, minimal.

Using the image in \autoref{fig:wavefunctions}(a), we determine the kinetic energy distribution of the filtered \ce{Li+} ions, shown in \autoref{fig:wavefunctions}(c), and then, by multiplying with (127~u~+~7~u)/127~u (see \autoref{eq:Ekin}), the distribution of the total kinetic energy release $P(E_\text{kin}^\text{tot})$. Finally, using the reflection approximation we obtain $P(R)$ from $P(E_\text{kin}^\text{tot})$ via $V_\text{QC}$ by applying a standard transformation of probability distributions

\begin{equation}
P(R) = P(E_\text{kin}^\text{tot})\left|\frac{\mathrm{d}E_\text{kin}^\text{tot}}{\mathrm{d}R}\right| = P(E_\text{kin}^\text{tot})\left|\frac{\mathrm{d}V_{QC}(R)}{\mathrm{d}R}\right|,
\end{equation}

as described previously~\cite{kristensen_laser-induced_2023}. The solid black curve in \autoref{fig:wavefunctions}(d) shows the resulting $P(R)$, denoted $P_\text{QC}(R)$, and the solid red curve shows the theoretical reference $P_\text{theo}(R)$. For $P_\text{QC}(R)$ the peak occurs at $R_\text{peak} = 2.4$~{\AA}, which matches the peak position $R_\text{peak} = 2.40$~{\AA} of $P_\text{theo}(R)$. It is seen that $P_\text{QC}(R)$ is broadened towards lower $R$-values compared to $P_\text{theo}(R)$, and the FWHM of $P_\text{QC}(R)$, 0.41~{\AA}, is about 52~\% larger than the FWHM, 0.27~{\AA}, of $P_\text{theo}(R)$. Varying the momentum criteria in \autoref{eq:filter_rad} with $\pm 7$~$\text{amu}\cdot\text{km} / \text{s}$ and the width of the angle criteria in \autoref{eq:filter_ang} with $\pm 10\degree$ changes the FWHM of $P_\text{QC}(R)$ with at most 0.02~{\AA}.

\Autoref{fig:LiI_potentials}(b) shows that in addition to the ground state there are 9 energetically-different excited states of \ce{LiI^{2+}}.  What happens if the laser-induced double ionization populates some of these excited states and thus \ce{LiI^{2+}} fragments via higher-lying potential curves? To explore this, we determined $P(R)$ as described above but now via potential curves for the excited states. First, we note that the $^3\Sigma_1^-$ ground state is split in two close-lying states (due to interaction with the $^3\Pi_0$ state) and that the internuclear distributions determined via these two states are essentially identical. In contrast, $P(R)$ determined via the $^3\Pi_2$ curve, lying above the two $^3\Sigma_1^-$ states and having a different curvature, peaks at $R$~=~2.6~{\AA} and has a width of 0.60~{\AA}, i.e., it deviates significantly from $P_\text{theo}(R)$. Similarly, when $P(R)$ is determined via the next potential curve, $^3\Pi_1$, we find an internuclear distribution with a peak at $R$~=~2.6~{\AA} and a width of 0.52~{\AA}, so again a significant deviation from $P_\text{theo}(R)$.

The good agreement between $P(R)$ determined via the $^3\Sigma_1^-$ ground state and $P_\text{theo}(R)$, \autoref{fig:wavefunctions}(d), indicates that the double ionization mainly populates \ce{LiI^{2+}} in this state and much less in the higher-lying excited states. To assess if this makes sense, we consider the double ionization process. The ionization energy, $I_p$, for \ce{LiI} and \ce{LiI+} is $\sim$~8~eV  and $\sim$~16~eV, respectively, see \autoref{fig:LiI_potentials}(a). The corresponding Keldysh parameters, $\sqrt{\frac{I_p}{2U_p}}$~\cite{posthumus_dynamics_2004}, are $\sim$~0.37 and $\sim$~0.54 ($U_p$~$\sim$~28~eV is the ponderomotive energy) and indicates ionization in the tunneling regime. At the intensity of the laser pulse, $4.7 \times 10^{14}$~W/cm$^2$, ionization of \ce{LiI} should be saturated and since $I_p$ for \ce{LiI+} is twice as large as $I_p$ for \ce{LiI}, we expect that double ionization occurs sequentially~\cite{posthumus_dynamics_2004}.

Now we note that theoretical modelling of tunnel ionization of many-electron atoms and molecules, induced by an intense, linearly polarized laser pulse, have shown that the ionization rate depends on the asymptotic form of the orbital, from which an electron is removed, near the tunnel exit~\cite{tolstikhin_weak-field_2014}. The electron configuration of LiI is $\sigma^2\pi_x^2\pi_y^2$. The $\sigma$-orbital is more localized than the $\pi$-orbitals, i.e., the $\sigma$-orbital has less density near the tunnel exit. With reference to the many-electron tunneling theory, we therefore expect that ionization of \ce{LiI} is most likely to occur due to removal of an electron from the $\pi$-orbital, and the electron configuration of \ce{LiI+} will thus be $\sigma^2\pi^3$. For the second ionization step, it is again most likely to remove an electron from a $\pi$-orbital and so the electron configuration of \ce{LiI^{2+}} will be  $\sigma^2\pi^2$.

The $\sigma^2\pi^2$ configuration gives rise to the terms $^3\Sigma_1^-$, $^1\Delta_2$ and $^1\Sigma_0^+$, see \autoref{tab:correlations} in the appendix. The former is the ground state while the two singlet states are lying $\sim$~1.6~eV and $\sim$~2.3~eV higher (at $R$ = 2.4 {\AA}). We determined $P(R)$ via each of the two singlet states. For $^1\Delta_2$, $P(R)$ is very close to that obtained via the ground state reflecting the fact that the two corresponding potential curves are very similar except for a vertical displacement. Using the potential curve for $^1\Sigma_0^+$, we find a $P(R)$ peaked at $R$~=~2.8~{\AA} and a width of 0.75~{\AA}. Since this state is $\sim$~2.3~eV higher than the ground state, we expect, however, that its probability of formation compared to that of the ground state is significantly reduced due to the strongly decreasing ionization rate of tunnel ionization when the ionization potential increases. For the same reason, formation of \ce{LiI^{2+}} in the $^1\Delta_2$ state should also be reduced compared to the ground state, noting that even if $^1\Delta_2$ is somewhat populated, it will not change $P(R)$ due to the similarity of the potential curve with that of the ground state.

The above considerations are qualitative, but they give an indication of the underlying reasons why double ionization mainly populates the $^3\Sigma_1^-$ ground state of \ce{LiI^{2+}}, i.e., the dependence of tunnel ionization on orbital structure and ionization potential.

As said, the peak of $P_\text{QC}(R)$ matches that of $P_\text{theo}(R)$ but $P_\text{QC}(R)$ is broadened towards higher $R$-values compared to $P_\text{theo}(R)$. We believe that the broadening results from internuclear motion during the laser pulse and enhanced ionization of \ce{LiI+}. The equilibrium distance, $R_\text{eq}$ of \ce{LiI}, 2.4 {\AA}, is less than that of \ce{LiI+}, 2.7 Å, see \autoref{fig:LiI_potentials}(a). Thus, when LiI is ionized and \ce{LiI+} is formed, likely in the ground state, there is a force acting on the nuclei tending to increase their distance. If there is a small delay before the second ionization happens, $R$ will increase a bit. Assuming that the delay, which must be (somewhat) smaller than the pulse duration, is 10 fs (15 fs), we find that $R$ increases by $\sim$~0.1 {\AA} ($\sim$~0.2 {\AA}) based on a classical simulation using the potential curve for the ground state of \ce{LiI+}. Increases of $R$ of this magnitude are consistent with the observed broadening of $P_\text{QC}(R)$.

An additional effect that could, in principle, broaden $P_\text{QC}(R)$ towards larger $R$-values relative to $P_\text{theo}(R)$ is thermal population in high rotational states, for which centrifugal distortion increases the effective internuclear separation. To assess this, we numerically calculated the distribution of a thermal rovibrational ensemble of LiI molecules at $T=430^\circ$C. The rotational degree of freedom was included by adding the centrifugal distortion term, $\hbar^2 J(J+1)/(2\mu R^2)$, in the Hamiltonian of the time-independent Schr\"{o}dinger equation, where $J$ is the rotational quantum number and $\mu$ is the reduced mass. At this temperature, the resulting modification to the $R$-distribution was negligible, showing that rotational excitation and the associated centrifugal distortion cannot account for the observed discrepancy.

In addition, enhanced ionization of \ce{LiI+} may also play a role. Here, enhanced ionization refers to the fact that the tunneling ionization rate of a molecular cation, in our case of \ce{LiI+}, depends on the internuclear distance~\cite{seideman_role_1995,constant_observation_1996}. In particular, for many diatomic molecules, the ionization rate increases when $R$ is increased beyond $R_\text{eq}$ to reach a maximum at a critical distance $R_\text{c}$~\cite{posthumus_dynamics_2004}. For instance for \ce{I2+} $R_\text{eq}$ = 2.64 {\AA}~\cite{zhao_spectroscopic_2024} while $R_\text{c}$ $\sim$~5 {\AA}~\cite{posthumus_field-ionization_1996}. We are not aware of any theoretical or experimental results for \ce{LiI+} but assuming it behaves similar to other diatomics, the ionization rate will increase as $R$ is increased in the region around and above $R_\text{eq}$ up to $R_\text{c}$ for this molecule. This $R$-dependence will tend to skew the internuclear distribution in a way that enhances the larger $R$-values, which is consistent with the experimental observations.

The effect of internuclear motion can be minimized and possibly effectively eliminated by employing a shorter laser pulse with a duration of about 10 fs. The effect of enhanced ionization would still represent a distortion of the internuclear distribution upon its projection on the \ce{LiI^{2+}} potential curve. To remove this effect, it would be necessary to calculate the $R$-dependence of the \ce{LiI+} ionization step and then deconvolve it from the experimental data, similar to that done in Ref.~\cite{zeller_determination_2018}. In summary of this section, the experimentally determined $P(R)$ match the theoretically determined $P(R)$ in terms of peak position but it is somewhat broadened towards $R$-values larger than $R_\text{eq}$. We believe the broadening is mainly due to a slight internuclear motion of the transiently populated \ce{LiI+} ion and its $R$-dependent ionization rate.

Finally, to demonstrate the shortcomings of using a pure Coulomb potential, we also determined $P(R)$ using \autoref{eq:Coulomb} directly.  The result, denoted $P_\text{Coul}(R)$, is represented by the dashed curve in \autoref{fig:wavefunctions}(d). The poor agreement with $P_\text{theo}(R)$ clearly illustrates the need for potential curves from high-level quantum chemistry calculations in order to convert the measured kinetic energy distribution into an accurate distribution of internuclear distances.

\subsection{LiI dimer}

\begin{figure}[t]
\includegraphics[width = 8.6cm]{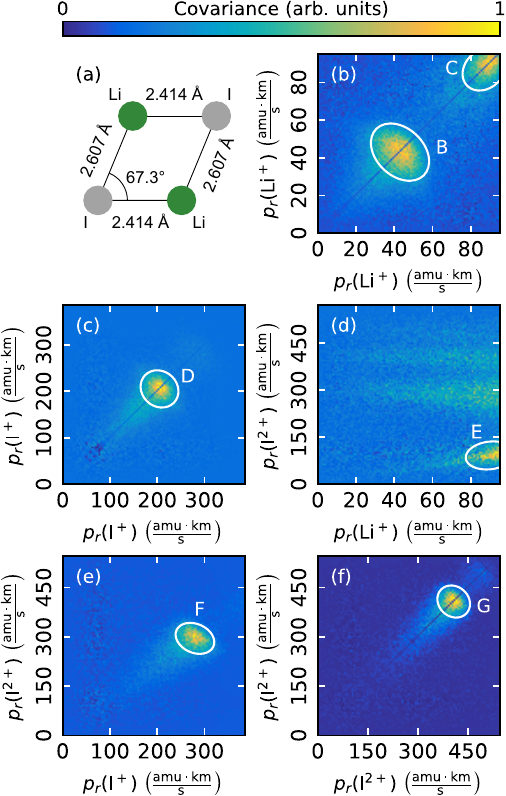}
\caption{(a) Sketch of the parallelogram-shaped, C$_{2h}$ symmetric, equilibrium structure of \ce{(LiI)2}~\cite{torring_structure_1996}. (b)--(f): Covariance maps of the radial momentum distributions for \ce{Li+}, \ce{I+}, and \ce{I^2+}, plotted with a linear color scale. The regions with significant signal are labeled from B to G.}
\label{fig:radial_covariances}
\end{figure}

So far, we discussed Coulomb explosion of \ce{LiI} into a pair of \ce{Li+} and \ce{I+} ions giving rise to the peaks in the respective $P(E_\text{kin})$ at 4.7~eV and 0.3~eV. As seen in \autoref{fig:images_distributions}(a4)--(b4), there are also other peaks in the kinetic energy distributions. Now we show that one of these originates from Coulomb explosion of \ce{LiI} into a pair of \ce{Li+} and \ce{I^{2+}} ions and the rest from various Coulomb explosion channels of the \ce{LiI} dimer, \ce{(LiI)_2}. The presence of the dimer in our target gas is expected since it is known that vapors of most alkalide halides contain a certain amount of dimers~\cite{hargittai_molecular_2000}. In fact, as mentioned in \autoref{sec:setup}, the dimer is likely more abundant than the monomer under our experimental conditions. A sketch of the equilibrium structure of \ce{(LiI)_2} is given in \autoref{fig:radial_covariances}(a)~\cite{hargittai_molecular_2000}. In the following, we also include data based on the 2D momentum image for \ce{I^{2+}} ions, \autoref{fig:images_distributions}(c1), that was also recorded. Similar to the momentum images for \ce{Li+} and \ce{I+}, the image exhibits several distinct half-rings that also stand out in the radial momentum distribution, \autoref{fig:images_distributions}(c3), and in the kinetic energy distribution, \autoref{fig:images_distributions}(c4).

To trace the source of the ions in the unassigned peaks in the three $E_\text{kin}$ distributions [\autoref{fig:images_distributions}(a4)--(c4)], we again use covariance analysis. \Autoref{fig:radial_covariances} shows the covariance maps determined for the different combinations of the radial momentum distributions of \ce{Li^+}, \ce{I^+}, and \ce{I^2+}, excluding the one that was already shown in \autoref{fig:Li_I_covariance}(a). Six distinct regions in the radial covariance maps are observed, and they are labeled B to G. To identify the origin of the fragment ions in these regions, we performed classical dynamics simulations of Coulomb explosion of \ce{(LiI)2} for various possible fragmentation channels. The starting point is the dimer in its equilibrium geometry, which is then suddenly ionized, i.e., electrons are removed instantaneously. The fragmentation dynamics of the resulting multiply-charged cationic dimer is hereafter obtained by calculating the trajectories of the neutral and charged fragments via the appropriate pair potentials determined from quantum chemistry calculations. In particular, we determined the pair potentials of (\ce{Li+}, I), (\ce{Li+}, \ce{I^{2+}}), (\ce{I}, \ce{I+}), (\ce{I}, \ce{I^{2+}}), (\ce{I+}, \ce{I+}), (\ce{I+}, \ce{I^{2+}}), and (\ce{I^{2+}}, \ce{I^{2+}}). The computational method and the level of the coupled-cluster theory were the same as for the (\ce{Li+}, \ce{I+}) pair potential, $V_\text{QC}$, described in \autoref{sec:principle} but without including spin-orbit interactions. All pair potentials except that for (\ce{Li+}, \ce{Li+}), which was treated as a pure Coulomb potential, are available in the supplementary information. We stress that the simulations are not expected to provide accurate descriptions of the Coulomb explosion processes since they do not account for, e.g., the initial ionization dynamics, charge transfer and possible structural deformation during the laser pulse. Additionally, the experiment only measures the projection of the emission angles of the ions on the detector plane whereas the simulations provide the full 3D angles, i.e., the experimental and simulated emission angles are not perfectly comparable. Yet, as discussed below the simulations are helpful for identifying the charge state of the dimer cation and which fragments it breaks into.

\begin{figure*}[t]
\includegraphics[scale = 1]{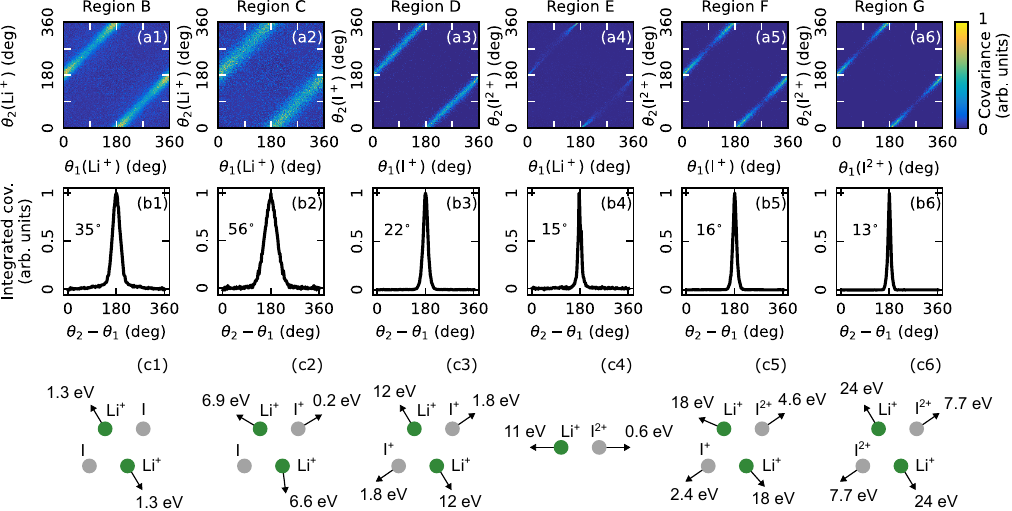}
\caption{(a1)--(a6): Covariance maps for the ion pairs in the regions defined in \autoref{fig:radial_covariances}, plotted with a linear color scale. (b1)--(b6): Integration of the covariance images in (a1)–(a6) along $\theta_1$ as a function of the difference angle $\theta_2 - \theta_1$~\cite{christiansen_laser-induced_2016}. The FWHM of the peaks are annotated next to the peaks. (c1)--(c6): Sketches of the fragmentation channels assigned as the origin for each of the six regions.}
\label{fig:angular_covariances}
\end{figure*}

First, region B in \autoref{fig:radial_covariances}(b) shows that pairs of \ce{Li+} ions with similar radial momenta $p_\text{r}\sim 46~\text{amu}\cdot\text{km} / \text{s}$, are correlated. This momentum corresponds to $E_\text{kin} \sim 1.6$~eV, which matches the peak centered at 1.6~eV in \autoref{fig:images_distributions}(a4). Since two correlated \ce{Li+} ions cannot come from the monomer, the ions in region B must originate from Coulomb explosion of the dimer. This is corroborated by the fact that the correlated \ce{Li+} ions are ejected with a relative angle of 180$\degree$, as shown by the angular covariance map calculated for the ions in region B, see \autoref{fig:angular_covariances}(a1)--(b1). For double ionization of \ce{(LiI)2} leading to fragmentation into two \ce{Li+} ions and two neutral \ce{I} atoms, \autoref{fig:angular_covariances}(c1), the simulations show that each \ce{Li+} ion obtains a final $E_\text{kin} \sim 1.33$~eV, which is fairly close to the experimental value of 1.6 eV. Furthermore, the covariance maps in \autoref{fig:Li_I_covariance}(a) and \autoref{fig:radial_covariances}(d) show that the \ce{Li+} ions in region B do not correlate with \ce{I+} or \ce{I^{2+}}. Therefore, we ascribe region B to double ionization of \ce{(LiI)2} and the subsequent fragmentation:  \ce{(LiI)2^{2+}}~$\rightarrow$~2\ce{Li+}~+~2\ce{I}.

Double ionization of the dimer could in principle also give rise to the following fragmentation channels: \ce{(LiI)2^{2+}}~$\rightarrow$~\ce{Li+}~+~\ce{I+}+~\ce{LiI} or \ce{(LiI)2^{2+}}~$\rightarrow$~\ce{Li+}~+~\ce{I+}+~\ce{Li}~+~\ce{I}. We briefly address them because \ce{Li+} and \ce{I+} ions produced in this manner would introduce some unwanted distortion in the determination of the internuclear distribution of the \ce{LiI} monomer. We do, however, think that the two channels are highly unlikely for the following reason. The charges on the \ce{Li} and \ce{I} atomic sites were determined using the MBIS (Minimal Basis Iterative Stockholder) decomposition of the electron density calculated with the $\omega$B97X-D functional and the uncontracted Sapporo-2012-AQZP basis set~\cite{verstraelen_minimal_2016,lu_multiwfn_2012,noro_segmented_2012}. We found 0.92~e, -0.92~e, respectively. Thus, in a simple picture, we can describe each \ce{LiI} monomer in the dimer as \ce{Li+-I-}. The interaction with the laser pulse will first remove the most loosely bound electrons, which are the ones from the two \ce{I-} with their electron affinity of 3.1 eV. This leads to 2\ce{Li+}~+~2\ce{I}, i.e., the \ce{Li+} ions in region B discussed above. To produce ~\ce{Li+}~+~\ce{I+} along with either \ce{LiI} or \ce{Li}~+~\ce{I} requires that two electrons are removed from one I-site and none from the other. Removal of the second electron from the I-site corresponds to ionization of \ce{I}, i.e., an ionization potential of 10.5 eV. It appears highly improbable that this should happen without the laser pulse detaching one electron from the other I-site (binding energy of 3.1 eV).

\begin{figure}
\includegraphics[width = 8.6cm]{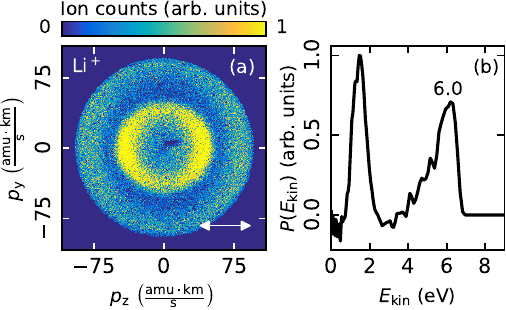}
\caption{(a) \ce{Li+} ion image after selecting the \ce{Li+} ions that correlate with another \ce{Li+} ion in the same laser shot and that also fulfill $\left| p_{\text{r}_1}(\text{Li}^+) - p_{\text{r}_2}(\text{Li}^+) \right| < 14$~$\frac{\text{amu}\cdot\text{km}}{\text{s}}$ and
$-40\degree < \left| \theta_1(\text{Li}^+)-\theta_2(\text{Li}^+) \right| - 180\degree < 40\degree$, plotted with a linear color scale. (b) Kinetic energy distribution for the \ce{Li+} ions in the image in (a). The ions from region C, which are otherwise obscured by ions from region A, are now revealed.}
\label{fig:filtering}
\end{figure}

For region C, \autoref{fig:radial_covariances}(b), we also observe correlations between two \ce{Li+} ions with the same momenta, $p_\text{r} \sim 90~\text{amu}\cdot\text{km} / \text{s}$, which is close to the maximum momentum that can be recorded with the VMI spectrometer. As such, region C lies close to the covariance map edge. There is, however, overlap with the \ce{Li+} ions correlating with \ce{I^{2+}} in region E, \autoref{fig:radial_covariances}(d), discussed below, and with the \ce{Li+} ions in channel A, \autoref{fig:Li_I_covariance}(a). We therefore employ coincidence filters similar to those in \autoref{eq:filter_rad} and \autoref{eq:filter_ang} to remove all \ce{Li+} ions that do not correlate with another \ce{Li+} ion within the same laser shot. \Autoref{fig:filtering}(a) displays the filtered \ce{Li+} ion image and \autoref{fig:filtering}(b) the corresponding kinetic energy distribution. The peak centred at 6.0~eV corresponds to $p_\text{r} \sim 90~\text{amu}\cdot\text{km} / \text{s}$. In comparison, the molecular dynamics simulations show that \ce{Li+} ions from triple ionization of \ce{(LiI)2} and fragmentation into 2\ce{Li+} + \ce{I+} + \ce{I}, \autoref{fig:angular_covariances}(c2), obtain $E_\text{kin}$ of 6.4~eV and 6.9~eV. This is quite close to the experimental $E_\text{kin}$ of 6.0~eV. Since no other Coulomb explosion pathway of \ce{(LiI)2} produces \ce{Li+} ions with a kinetic energy near 6.0~eV, we conclude that triple ionization and subsequent Coulomb explosion is responsible for region C. For completeness, we note that the simulated value of $E_\text{kin}$  for the \ce{I+} fragment is 0.2~eV. This value is consistent with a location of a peak on the side of, but not resolved from, the major peak in the experimental $P(E_\text{kin})$  produced by \ce{I+} ions from the \ce{(LiI)^{2+}}~$\rightarrow$~\ce{Li+}~+~\ce{I+} Coulomb explosion channel. Finally, the covariance map of the angular distribution of the \ce{Li+} ions in region C is centered at $180\degree$, \autoref{fig:angular_covariances}(a2)--(b2), whereas the simulations give an emission angle of $125\degree$. The reasons we cannot detect this molecular frame emission angle of the fragments ions are, we believe, the expected dependence of the strong-field triple ionization rate of the dimer on its spatial orientation and the projection of the 3D momentum of the recoil ions on the detector plane.  These limitations could likely be overcome by aligning the dimers prior to the arrival of the ionization pulse as demonstrated previously~\cite{hansen_control_2012,pickering_alignment_2018}, but it is much beyond the scope of the current work.

Region D shows that there are correlations between two \ce{I+} ions with the same momenta, $p_\text{r} \sim 210~\text{amu}\cdot\text{km} / \text{s}$, see \autoref{fig:radial_covariances}(c). Thus, these ions must originate from the dimer. The corresponding kinetic energy is 1.9~eV, see \autoref{fig:images_distributions}(b4). As a reference, we simulated the Coulomb explosion of a quadruply-ionized dimer, \ce{(LiI)2^{4+}}, into two \ce{Li+} and two \ce{I+} ions, \autoref{fig:angular_covariances}(c3). In this channel, each \ce{I+} ends with an $E_\text{kin}$ of 1.8~eV. The good match with the experimental value and, again, the fact that no other Coulomb explosion channel produces \ce{I+} ions with this kinetic energy, lead us to assign region D to the charge-symmetric Coulomb explosion following quadruple ionization of \ce{(LiI)2}. The sharp $180\degree$ correlation in the emission direction of the \ce{I+} ions, \autoref{fig:angular_covariances}(a3)--(b3), is in line with the simulated results and corroborates the assignment. The simulations show that the \ce{Li+} ions get a kinetic energy of 12~eV, which is beyond what can be detected with the current experimental setup. Therefore, we do not expect to see correlations between \ce{Li+} and \ce{I+} from the quadruple ionization channel, consistent with our observations in the radial covariance analysis, \autoref{fig:Li_I_covariance}(a). We point out that a similar Coulomb fragmentation process into four singly charged atomic fragment ions was reported recently for the HCl dimer~\cite{zhao_four-body_2025}.

Concerning region E, the covariance map in \autoref{fig:radial_covariances}(d) shows that \ce{Li+} and \ce{I^{2+}} with $p_\text{r} \sim 95~\text{amu}\cdot\text{km} / \text{s}$ are correlated. For the \ce{Li+} ions, this is the maximum momentum that can be recorded with the VMI spectrometer settings and, consequently, region E lies at the edge of the covariance map along the $p_\text{r}(\text{Li}^+)$ axis. The corresponding angular covariance map of these ions, \autoref{fig:angular_covariances}(a4), is very similar to the one for channel A [\ce{(LiI)^{2+}}~$\rightarrow$~\ce{Li+}~+~\ce{I+}]. This indicates that region E originates from triple ionization of \ce{LiI} and subsequent Coulomb explosion into \ce{Li+} + \ce{I^{2+}}. Our simulations show that such \ce{I^2+} ions obtain $E_\text{kin} \sim$~0.6~eV, which agrees with the peak in the kinetic energy distribution centred at 0.5~eV, \autoref{fig:images_distributions}(c4). For the \ce{Li+} ions, the simulated value of $E_\text{kin}$ is 11~eV, i.e. significantly more than the detected $E_\text{kin}$ of $\sim$~6.7 eV. The reason for this discrepancy is, we believe, that the ions in region E comes from those \ce{LiI} dimers that were sufficiently aligned out of the detector plane that the projection of the \ce{Li+} recoil ion momentum on the detector plane becomes small enough, i.e. around $95~\text{amu}\cdot\text{km} / \text{s}$, that the \ce{Li+} ion hits within the detector area. As such, region E represents only the tail of the signal stemming from the \ce{(LiI)^{3+}}~$\rightarrow$~\ce{Li+}~+~\ce{I^{2+}} channel.

The last two regions, F and G, concern \ce{I^{2+}} ions. \Autoref{fig:radial_covariances}(e) show correlations between \ce{I+} ions with $p_\text{r} \sim 290~\text{amu}\cdot\text{km} / \text{s}$ and \ce{I^{2+}} ions with $p_\text{r} \sim 305~\text{amu}\cdot\text{km} / \text{s}$. The corresponding  $E_\text{kin}$ is 3.4~eV for \ce{I+} and 3.7~eV for \ce{I^{2+}} see \autoref{fig:images_distributions}(b4)--(c4). Since we already identified double, triple and quadruple ionization of the dimer, we expect the next channel in the progression is quintuple ionization and fragmentation into \ce{I^{2+}}~+~\ce{I+}~+~2\ce{Li+}, \autoref{fig:angular_covariances}(c5). We simulated this process and found that $E_\text{kin}$ of \ce{I+} is 2.4~eV, and 4.6~eV for \ce{I^{2+}}. While the agreement with the experimental values of 3.4~eV for \ce{I+} and 3.7~eV for \ce{I^{2+}} is not perfect the simulated values are close enough that we assign channel F to \ce{LiI^{5+}}~$\rightarrow$~\ce{I^{2+}}~+~\ce{I+}~+~2\ce{Li+}, noting that no other Coulomb fragmentation channel can produce a pair of correlated \ce{I^{2+}} and \ce{I+}.

Finally region G shows the correlation between two \ce{I^{2+}} ions with $p_\text{r} \sim 405~\text{amu}\cdot\text{km} / \text{s}$  corresponding to $E_\text{kin} \sim$~6.9~eV, see \autoref{fig:images_distributions}(c4). The simulation of sixtuple ionization and fragmentation into two \ce{Li+} ions and two \ce{I^{2+}} ions shows that for the latter $E_\text{kin} \sim$~7.7~eV. The match between the experimental value of 6.9~eV for \ce{I^{2+}} and the simulated values is good and supports \ce{(LiI)2^{6+}}~$\rightarrow$~2\ce{Li+}~+~2\ce{I^{2+}} as the origin of region G. For this charge-symmetric explosion channel we see a sharp $180\degree$ relative emission of the \ce{I^2+} ions similar to that observed for the \ce{I+} ions from the quadruple ionization channel. For completeness, we note that the simulated $E_\text{kin}$ of the \ce{Li+} ions is $\sim$~24~eV, so most of the ions end outside the detector, hindering a proper analysis with the current dataset and statistics.

\section{Conclusion and outlook}\label{sec:conclusion}

The focus of this work was femtosecond laser-induced Coulomb explosion of gas-phase \ce{LiI} molecules and \ce{LiI} dimers. The key result for the \ce{LiI} monomer is that the kinetic energy distribution $P(E_\text{kin})$ of the \ce{Li+} fragments, produced when \ce{LiI} is doubly ionized and subsequently fragments into a (\ce{Li+}, \ce{I+}) ion pair, can be used to determine the distribution of internuclear distances $P(R)$ of \ce{LiI} and that this distribution agrees well with the theoretically expected $P(R)$. The conversion of $P(E_\text{kin})$ to $P(R)$ is done via the potential energy curve for the $^3\mathrm{\Sigma}_1^-$ ground state of \ce{LiI^{2+}} obtained by an \emph{ab initio} calculation. We discussed why the population of excited states of \ce{LiI^{2+}} is suppressed due to the dependence of tunnel ionization on orbital structure and on the energy of the excited states. As such, the fragmentation of \ce{LiI^{2+}} appear to mainly occur via the ground state, which ensures the one-to-one correspondence between $E_\text{kin}$ and $R$ needed to retrieve $P(R)$.

The molecular sample also contains \ce{LiI} dimers with an expected abundance twice that of the \ce{LiI}  monomers. Therefore, it was necessary to use coincidence filtering to extract the \ce{Li+} and \ce{I+} ions originating from Colomb explosion of the \ce{LiI} monomer.  Concerning the dimer, we used covariance analysis of the radial momentum distributions and of the angular distributions of the \ce{I+}, \ce{Li+} and \ce{I^{2+}} fragments to show that the laser pulse ionizes it from two up to six times and that each multiply-charged dimer cation leads to a characteristic Coulomb fragmentation pattern. For instance, the quadruply-charged \ce{LiI} dimer cation, \ce{(LiI)2^{4+}} explodes into two \ce{Li+} and two \ce{I+} recoil ions.

We point to the following opportunities opened by our work. Photodissociation of alkali halides involves the coupling between potential curves with predominantly covalent and ionic character. More than 35 years ago, Zewail and coworkers illuminated the photodissociation dynamics of alkali halide molecules, notable of \ce{NaI}, in pioneering studies employing femtosecond laser-induced fluorescence spectroscopy~\cite{rose_femtosecond_1989}. Now, time-resolved Coulomb explosion should make it possible to directly image the time-dependent internuclear distributions, as the vibrational wave packet induced by a fs pump pulse evolves and splits between the bound and dissociative parts of the potential curves. As a first step, it will be interesting and relevant to apply laser-induced Coulomb explosion to other alkali halides and determine the internuclear distributions of the static molecules.  Regarding the \ce{LiI} dimer, the current work only aimed at identifying its presence and the different Coulomb explosion channels produced by the laser interaction. The dimer could, however, also be an interesting target for observing structural dynamics. For instance, in the Coulomb explosion resulting from the double ionization, \autoref{fig:angular_covariances}(c1), or from quadruple ionization, \autoref{fig:angular_covariances}(c3), the light \ce{Li+} ions fly away very quickly. It should then be possible to measure the time-dependent internuclear separation of the two \ce{I} atoms or two \ce{I+} ions by further ionizing them with a more intense optical probe pulse or, perhaps better, with an x-ray pulse from a free-electron laser that resonantly knocks off an inner-shell electron and induces multiple ionization~\cite{young_femtosecond_2010,rorig_multiple-core-hole_2023}.

\begin{acknowledgments}
H.S. acknowledges support from The Villum Foundation through a Villum Investigator Grant No. 25886. We thank Jan Thøgersen for expert help on the optics and the laser system and Julie Olsen for making the optical setup that produced the 25 fs laser pulses.
\end{acknowledgments}

\appendix*

\section*{Appendix: Molecular states of L\lowercase{i}I$^{2+}$}\label{sec:Appendix}

There are two sets of valence orbitals: 2s and 2p for \ce{Li+} and 5s and 5p for \ce{I+}, with the lowest electron configuration is 5s$^2$5p$^4$2s$^0$2p$^0$. The Li orbitals are significantly higher in energy and are not important. The \ce{LiI^{2+}} states can thus be considered as \ce{I+} states perturbed by \ce{Li+}.

The level of calculation is (6,4)-CASSCF followed by NEVPT2~\cite{guo_approximations_2021} and inclusion of spin-orbit in the mean-field approximation~\cite{neese_efficient_2005}, as implemented in the ORCA program package~\cite{neese_software_2022}. The (6,4) indicates that the six electrons are distributed in all possible ways in four (5s and 5p) orbitals. Scalar relativistic effects are not included. We extract three triplet states and six singlet states, using equal state averaging over the triplet and singlet blocks, and equal weighting of the three and six states within each block. Alternative weighting gives similar results, but some alternative weighting schemes may break symmetry. The basis sets are the Sapporo-XZP-2012 (X=D,T,Q)~\cite{noro_segmented_2012} with an L$^{-3}$ basis set extrapolation of the X=T,Q results.

Within C$_{\infty v}$, the s- and $\text{p}_\text{z}$-orbitals are $\sigma$, while the $\text{p}_\text{x}$ and $\text{p}_\text{y}$ are $\pi$. The 5s-orbital is always doubly occupied, but the four electrons in the 5p orbitals can be distributed in several different ways. The possible different electron configurations are given in~\autoref{tab:electron_config}. The first three (Labels 1-3) are open shells and each have four microstates, leading to one singlet and one triplet state, while the last three only have doubly occupied orbitals and must be singlets. There are thus nine triplet and six singlet microstates. Labels 2 and 3, and 4 and 5, are degenerate, and there are therefore two triplet and four singlet energetically different states. This is confirmed at the CASSCF and NEVPT2 level. The representation product in C$_{\infty v}$ indicates $\Sigma \times \Pi = \Pi$ and $\Pi \times \Pi = \Sigma^+ + [\Sigma^-] + \Delta$, i.e., the triplet can only be $\Sigma^-$ while the singlets can only be $\Sigma^+$ and $\Delta$. This gives the term symbols in \autoref{tab:electron_config}. The two triplet states with $\Sigma$ and $\Pi$ symmetry have degeneracies 3 and 6, respectively, and the four singlet states with $\Sigma$, $\Delta$, $\Pi$, and $\Sigma$ symmetry have degeneracies 1,2,2, and 1, respectively. In the atomic limit, these must correspond to $^3$P, $^1$D, and $^1$S states, which agrees with degeneracies of 9, 5 and 1.

\begin{table}[]
\centering
\caption{Orbital occupancies and resulting term symbols.}\label{tab:electron_config}
\begin{tabular}{lllll}
%\arrayrulecolor{black}
Label  & \multicolumn{2}{l}{Microstates} & \multicolumn{2}{l}{Term} \\ \hline
       & Singlet        & Triplet        & Singlet     & Triplet    \\ \cline{2-5}
1 \quad $\sigma^2\pi_x^1\pi_y^1$ & 1 & 3 & $^1\Sigma^+$, $^1\Delta$  & $^3\Sigma^-$ \\
2 \quad $\sigma^1\pi_x^2\pi_y^1$ & 1 & 3 & $^1\Pi$                    & $^3\Pi$      \\
3 \quad $\sigma^1\pi_x^1\pi_y^2$ & 1 & 3 & $^1\Pi$                    & $^3\Pi$      \\
4 \quad $\sigma^2\pi_x^2\pi_y^0$ & 1 & 0 & $^1\Sigma^+$, $^1\Delta$   &              \\
5 \quad $\sigma^2\pi_x^0\pi_y^2$ & 1 & 0 & $^1\Sigma^+$, $^1\Delta$   &              \\
6 \quad $\sigma^0\pi_x^2\pi_y^2$ & 1 & 0 & $^1\Sigma^+$               &
\end{tabular}
\end{table}

When including spin-orbit, the atomic terms become: $^3$P$_2$, $^3$P$_1$, and $^3$P$_0$ with degeneracies 5, 3, and 1 and $^1$D$_2$ and $^1$S$_0$ with degeneracies 5 and 1, giving nine triplet and six singlet microstates. \Autoref{tab:energy_comparison} compares the calculated energies to experimental values from the literature~\cite{martin_spectrum_1960}. The values are in decent agreement, given that scalar relativistic effects are ignored. The calculations confirm the degeneracies 5, 3, 1, 5, and 1.

In the molecular case, the terms become: $^3\Sigma_1^-$, $^3\Pi_2$, $^3\Pi_1$, $^3\Pi_0$ $^1\Delta_2$, and $^1\Sigma_0^+$, see \autoref{tab:correlations}. For the singlet states, the energetic order is: $^1\Delta_2$, $^1\Pi_1$, $^1\Sigma_0^+$, and $^1\Sigma_0^+$ with degeneracies 2, 2, 1, and 1. For the triplet states, the energetic order is: $^3\Sigma_1^-$, $^3\Pi_2$, $^3\Pi_1$, and $^3\Pi_0$, with expected degeneracies 3, 2, 2, and 2, however, the interaction of the $M_j$ = 0 components of $^3\Sigma_1^-$ and  $^3\Pi_0$ leads to degeneracies of 1, 2, 2, 2, 1, and 1.

\begin{table}[]
\centering
\caption{Comparison between calculated energies for the \ce{I+} states and experimental values from the literature~\cite{martin_spectrum_1960}.}\label{tab:energy_comparison}
\begin{tabular}{llll}
State & \multicolumn{2}{l}{Energy (cm$^{-1}$)}        &        \\ \hline
      & Exp.            & \multicolumn{2}{l}{Calc.}  \\ \cline{2-4}
$^3\mathrm{P}_2$   & 0              & \multicolumn{2}{l}{0}     \\
$^3\mathrm{P}_1$   & 7087           & \multicolumn{2}{l}{5840}  \\
$^3\mathrm{P}_0$   & 6448           & \multicolumn{2}{l}{6172}  \\
$^1\mathrm{D}_2$   & 13727          & \multicolumn{2}{l}{13260} \\
$^1\mathrm{S}_0$   & 29501          & \multicolumn{2}{l}{30058}
\end{tabular}
\end{table}

\begin{table}[]
\centering
\caption{Term symbols when including spin-orbit effects. $\#$ indicates number of energetically different states.}\label{tab:correlations}
\begin{tabular}{lllll}
\multicolumn{2}{l}{\begin{tabular}[c]{@{}l@{}}Dominant electron\\ configs. \end{tabular}} & Sym C$_{\infty v}$ & \# & Atomic terms \\ \hline
$\sigma^2\pi_x^1\pi_y^1$  & & $^3\Sigma_1^-$ & 1,2  & $^3$P$_2$  \\
$\sigma^1\pi_x^2\pi_y^1$  & $\sigma^1\pi_x^1\pi_y^2$ & $^3\Pi_2$ & 2 & $^3$P$_2$  \\
$\sigma^1\pi_x^2\pi_y^1$  & $\sigma^1\pi_x^1\pi_y^2$ & $^3\Pi_1$ & 2 & $^3$P$_1$  \\
$\sigma^1\pi_x^2\pi_y^1$  & $\sigma^1\pi_x^1\pi_y^2$ & $^3\Pi_0$ & 1,1 & $^3$P$_1$ + $^3$P$_0$ \\
$\sigma^2\pi_x^2\pi_y^0$  & $\sigma^2\pi_x^0\pi_y^2$ & $^1\Delta_2$  & 2& $^1$D$_2$   \\
$\sigma^1\pi_x^2\pi_y^1$  & $\sigma^1\pi_x^1\pi_y^2$ & $^1\Pi_1$ & 2 & $^1$D$_2$      \\
$\sigma^2\pi_x^1\pi_y^1$  & & $^1\Sigma_0^+$ & 1 & $^1$D$_2$   \\
$\sigma^0\pi_x^2\pi_y^2$  & & $^1\Sigma_0^+$ & 1 & $^1$S$_0$
\end{tabular}
\end{table}

%\bibliography{bibliography}


\end{document}